\documentclass[epj]{svjour}

\usepackage{graphicx,float}
\usepackage{booktabs,tabularx}

\begin{document}
\title{Suppression of dynamics and frequency synchronization in coupled slow and fast dynamical systems}
\subtitle{}
\author{Kajari Gupta \and G.Ambika
\thanks{\emph g.ambika@iiserpune.ac.in}%
}                     
\offprints{}          
\institute{Indian Institute of Science Education and Research, Pune 411008,India}
\date{\today}
%
\abstract{
We present our study on the emergent states of two interacting nonlinear systems with differing dynamical time scales. We find that the inability of the interacting systems to fall in step leads to difference in phase as well as change in amplitude. If the mismatch is small, the systems settle to a frequency synchronized state with constant phase difference. But as mismatch in time scale increases, the systems have to compromise to a state of no oscillations. We illustrate this for standard nonlinear systems and identify the regions of quenched dynamics in the parameter plane. The transition curves to this state are studied analytically and confirmed by direct numerical simulations. As an important special case, we revisit the well-known model of coupled ocean atmosphere system used in climate studies for the interactive dynamics of a fast oscillating atmosphere and slowly changing ocean. Our study in this context indicates occurrence of multi stable periodic states and steady states of convection coexisting in the system, with a complex basin structure. 
} 
\authorrunning{K.G et al}
\titlerunning{coupled slow and fast systems}
\maketitle
\section{Introduction}
\label{intro}
The complexity of several dynamical phenomena that occur in many physical, chemical, biological, geophysical and social systems arise due to the interaction of dynamical processes at differing time scales\cite{{free80},{paul04},{hansen11},{henzler07},{kiebel08},{break05},{anne11}}. For example fast and slow dynamical processes occur in modulated lasers and chemical reactions\cite{{das13},{zunino11}}. It is known that in biological processes, dynamics at time scales of days coexist and interact with dynamics of biochemical reactions in sub second time scales\cite{{mit02}}. There are several intracellular processes of differing time scales that directly or indirectly influence the electrical activity of neurons\cite{kay03}. On a global scale the weather and climate systems of earth contain subsystems spanning over widely differing time scales. Some of these subsystems are basically nonlinear and are strongly coupled like tropical atmospheric ocean system \cite{{dav91},{lind08}}. In this context some of the relevant questions addressed are how the fast dynamics can affect the predictability of the slow dynamics and how the slow and fast modes can be separated\cite{pe04}. In general, coupled slow and fast systems occur in engineering design where issues related to regulation and optimal control are relevant topics for study\cite{zvi10}. The method of analysis mostly followed in such contexts is adiabatic elimination of fast variables\cite{Sanders85} which is applicable when the time scales are widely different. Some of the recent studies clearly indicate that fast time scales can affect the slow dynamics in systems of different time scales\cite{{krupa08},{hid13}}. Therefore, a detailed study on possible emergent dynamics that occur due to interplay of time scales at all ranges is highly relevant and will have several applications.

In the context of coupled dynamical systems, some of the well studied emergent phenomena are synchronization, desynchronization, oscillation death, amplitude death(AD), phase locking etc.\cite{{kurths13},{arenas08},{saxena12},{valla02},{landa13},{Sakaguchi},{Ermentrout}}. The occurrence of amplitude death is reported in coupled oscillators as due to various mechanisms  like  dynamic coupling\cite{kon03}, time delay coupling\cite{ram98}, nonlinear coupling\cite{awa10}, conjugate coupling or coupling via dissimilar variables in identical oscillators, parameter mismatch and distributed frequencies\cite{{Yamaguchi},{Shiino},{Mirollo}} in coupling systems\cite{kar07}. Recently indirect coupling through an external environment is also reported as a general mechanism for AD\cite{{res11},{snehal}}. In all these studies the interacting systems are considered as evolving with the same dynamical time scale. We observe that in many real world systems, like the cases mentioned above, interacting systems can evolve with dissimilar time scales.  One or two studies in this direction have been recently reported, on populations with time scale diversity\cite{sil03} and spatio-temporal chaos with cascade of bifurcations caused by interactions among different time scales\cite{fuji03}. We note that there are many interesting open questions still to be addressed regarding the possible emergent dynamics, its characterization and transitions in coupled nonlinear systems with differing time scales.

In this work, we present the study on the dynamics of two coupled nonlinear systems that evolve with differing time scales such that one of them evolves slower than the other.  Our results indicate that for sufficient mismatch in time scale two such systems go into a synchronized state of amplitude death.  If the mismatch is small, with strong coupling the two systems get locked into a state of  frequency synchronization with a constant phase shift.  The resultant frequency as well as amplitude of the emergent state decreases as they approach amplitude death. In the weak coupling limit, we observe states of multi frequency followed by drift. We analyze the stability of the amplitude death state and the nature of transitions to this state as the parameters are tuned.  We present the results for standard nonlinear systems like coupled Landau-Stuart oscillators, R{\"o}ssler and Lorenz systems. As a specific case of relevance for coupled slow and fast systems, we consider the coupled ocean atmosphere model in climate studies and analyse its possible dynamical states.

\section{Coupled slow and fast systems}
We construct a simple model of two interacting systems with slow and fast dynamics by considering two identical systems evolving at different time scales. The equations governing their dynamics can in general be written as given below.

\label{sec:1}
\begin{flushleft}
\begin{eqnarray}
\dot{\bf {{X}_1}} = \tau _1{\bf {F(X_1)}} +\tau_1 \epsilon {\bf G H(X_1,X_2)}
\nonumber\\
\dot{\bf {{X}_2}} = \tau _2{\bf {F(X_2)}} +\tau_2 \epsilon {\bf G H(X_2,X_1)}
\label{gensf}
\end{eqnarray}
\end{flushleft} 
Here $ {\bf X_{1,2}}  {\bf \in} {\bf { R}^n} $, {\bf F} is the intrinsic dynamics of the system, {\bf H} denotes the coupling function, $ \epsilon $ the coupling strength. {\bf G } is an  n x n  matrix which decides the variables to be coupled. The parameters $ \tau_1 $ and $ \tau_2 $ decide the difference in time scales. Without loss of generality, we can take $ \tau_1=\tau $ and $ \tau_2 =1 $ with $ \tau $ as the time scale parameter, by tuning which, the time scale mismatch between the two systems can be varied. In this case, in addition to the coupling strength $ \epsilon $, the time scale mismatch parameter $ \tau $ also controls the asymptotic dynamics of the coupled systems.

\subsection{Coupled slow and fast periodic oscillators}
In this section we consider the case of two coupled periodic systems with differing time scales. As an example of a periodic oscillator, we take two Landau-Stuart oscillators with slow and fast time scales, coupled diffusively. Their dynamics then evolves as

\begin{eqnarray}
 \dot{x}_1 &=& \tau(((a-{x_1}^{2}-{y_1}^2)x_1-\omega y_1) + \tau\epsilon (x_2-x_1))
\nonumber \\
\dot{y}_1  &=&\tau((a-{x_1}^2-{y_1}^2)y_1+\omega x_1)
\nonumber \\
\dot{x}_2 &=&(a-{x_2}^2-{y_2}^2)x_2-\omega y_2 + \epsilon (x_1-x_2)
\nonumber \\
\dot{y}_2 &=&(a-{x_2}^2-{y_2}^2)y_2+\omega x_2
\label{landaueqn} 
\end{eqnarray}
 
\begin{figure}[H]

  \includegraphics[width = \columnwidth]{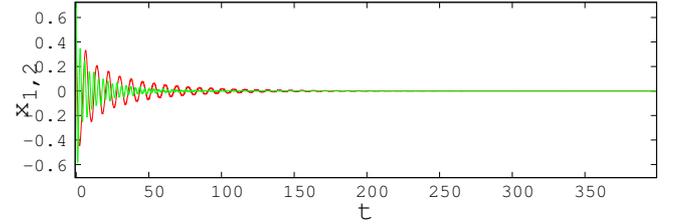}

\caption{(colour online) Time series of two coupled slow (red) and fast (green) Landau-Stuart oscillators in ~(\ref{landaueqn}) showing amplitude death for $ \tau = 0.4 $ and $ \epsilon = 0.3 $ .}
\label{ad1}       
\end{figure}

For $ a >0 $ the intrinsic oscillator has a limit cycle behaviour, and for $ a<0 $, a fixed point state. With $ a=0.1 $ and $ \omega=2 $ we analyse the system numerically using equation~(\ref{landaueqn}). We find that, for sufficiently large value of $ \epsilon $ and small value of $ \tau $, the two systems go into a state of amplitude death. This is shown in Fig~\ref{ad1} where the time series for the x- variable of both systems are plotted for $ \tau = 0.4 $ and $ \epsilon = 0.3 $.

\subsubsection{Stability analysis}
The systems in equation~(\ref{landaueqn}) go to a state of amplitude death when the fixed point of the whole system stabilizes. The parameters for which this happens can be obtained analytically by a linear stability analysis of the system about the fixed point. We find the fixed points of the systems in equation~(\ref{landaueqn}) are synchronized with $ (x^*,y^*) $ equal to $ (0,0) $. The stability of this state is decided by the eigenvalues of the Jacobian of the coupled slow and fast systems at this point given by,
\begin{equation}
 J =   \left (\begin{array}{cccccc}
\tau(a-\epsilon) & -\tau \omega & \tau \epsilon & 0 \\
\tau \omega & \tau a & 0 & 0 \\

\epsilon & 0  & a-\epsilon & -\omega  \\
0 & 0  & \omega & a  \end{array} \right) 
\label{jacobian}
\end{equation}

\begin{figure}[H]

  \includegraphics[width = \columnwidth]{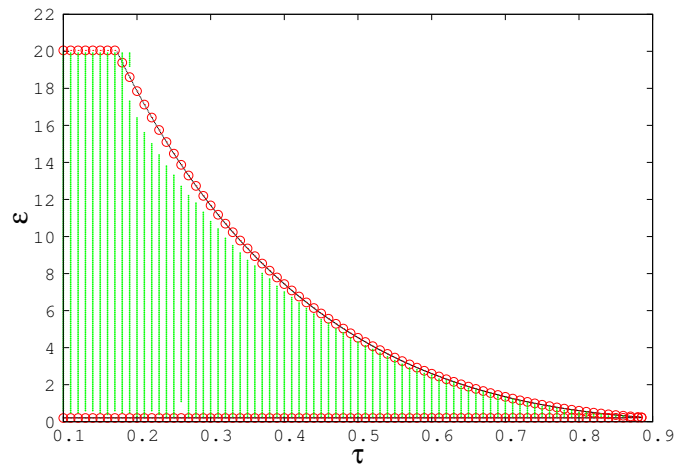}

\caption{(colour online) Amplitude death region for coupled slow and fast Landau-Stuart oscillators in the plane ($ \tau,\epsilon $). The green region corresponds to parameter values for AD obtained numerically while the black curve correspond to the transitions to AD obtained from stability analysis and the red circles are transition obtained from Routh-Hurwitz criterion.}
\label{paraplane}       
\end{figure}

The characteristic equation of the Jacobian is a 4th order polynomial of the form
\begin{eqnarray}
a_0\lambda ^4+a_1 \lambda^3+a_2 \lambda^2+a_3 \lambda +a_4 = 0
\label{characeqn}
\end{eqnarray}
where

\begin{eqnarray}
a_0&=&1 \nonumber \\
a_1&=&-2\tau a-2a+\tau \epsilon+\epsilon 
\nonumber \\
a_2&=&\tau^2 a^2+4\tau a^2-4\tau a\epsilon-\tau^2 \epsilon a+a^2 -\epsilon a+\omega^2 +\tau^2\omega^2 
\nonumber \\
a_3&=&-2\tau^2 a^3+3\tau^2 a^2\epsilon-2\tau a^3+3\tau a^2\epsilon -2\tau a\omega^2+\tau\epsilon\omega^2 
\nonumber \\
&&-2a\tau^2\omega^2+\tau^2\omega^2\epsilon 
\nonumber \\
a_4&=&\tau^2a^4-2\tau^2a^3\epsilon+2\tau^2a^2\omega^2-2\tau^2\omega^2a\epsilon+\tau^2\omega^4
\nonumber \\
\label{coeff}
\end{eqnarray}
From Routh-Hurwitz stability criterion\cite{routh}, the solutions for the eigenvalue $\lambda $ will have negative real parts if $a_i>0$,  $ \forall$ $ i $ and,
\begin{eqnarray}
Det   \left (\begin{array}{cccc}
a_1 & a_0  \\
a_3 & a_2  \end{array} \right)>0, \nonumber
   Det   \left (\begin{array}{cccccc}
a_1 & a_0 & 0  \\
a_3 & a_2 & a_1 \\
0 & a_4 & a_3 \end{array} \right)>0 \nonumber \\ 
Det   \left (\begin{array}{cccccccc}
a_1 & a_0 & 0 & 0  \\
a_3 & a_2 & a_1 & a_0 \\
0 & a_4 & a_3 & a_2 \\
0 & 0 & 0 & a_4 \end{array} \right)>0
\label{j}
\end{eqnarray}
Hence
\begin{eqnarray}
&&a_1a_2-a_0a_3>0 \nonumber \\
&&a_1a_2a_3-{a_1}^2a_4-a_0{a_3}^2 >0 \nonumber \\
&&a_1a_2a_3a_4-{a_1}^2{a_4}^2-a_0{a_3}^2a_4>0
\end{eqnarray}
The region in $\tau$ and $\epsilon$ plane where all of the above three conditions are satisfied is marked by the boundary line in blue in Fig.~\ref{paraplane}.  This thus identifies the region of amplitude death where the steady state of the coupled system is stable.

We directly evaluate the eigen values of J for different values of $ \tau $ and $ \epsilon $ and the values at which at least one of the eigen values changes from negative to positive are plotted to get the transition curves, shown in red in Fig~\ref{paraplane}. This boundary matches with that obtained by applying Routh-Hurwitz criteria directly.

We also do a detailed numerical analysis of the system in equation~(\ref{landaueqn}) for different values of these parameters scanning the parameter plane ($\tau,\epsilon $). To identify region of amplitude death in this plane, we compute the index $ A $ by taking the difference between global maximum and global minimum of the variable $x$ for each system after neglecting the transients. Then $A = 0 $ would indicate regions of AD\cite{res11}.
Using this we find the island of amplitude death in the ($\tau,\epsilon $) plane where both systems are found to settle to a synchronized fixed point, shown in green(Fig~\ref{paraplane}). It is clear that this region identified by direct numerical simulations agrees with that marked by the transition curves obtained from eigenvalues of the Jacobian.
\subsubsection {Frequency synchronization with a phase shift under strong coupling}
\begin{figure}[H]

  \includegraphics [width = \columnwidth]{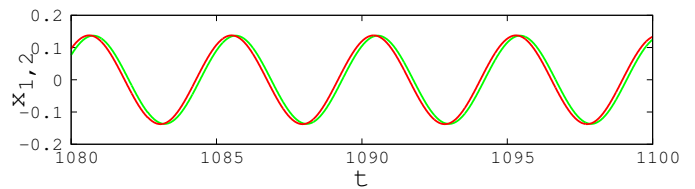}

\caption{(colour online)Time series of coupled slow (red) and fast (green) Landau-Stuart oscillator with $ \tau=0.4 $ and $\epsilon=10 $ indicating frequency syncronised state, phase shift.}
\label{ssff}       
\end{figure}

\begin{figure}[H]
\begin{center}

  \includegraphics [width = 0.5\columnwidth]{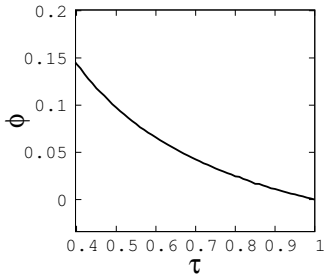}
\end{center}

\caption{(colour online)Phase shift between slow and fast Landau-Stuart oscillators $\phi$ as $\tau$ is varied with $\epsilon=10$.}
\label{st}       
\end{figure}

\begin{figure}[H]

 \includegraphics [width = 0.52\columnwidth]{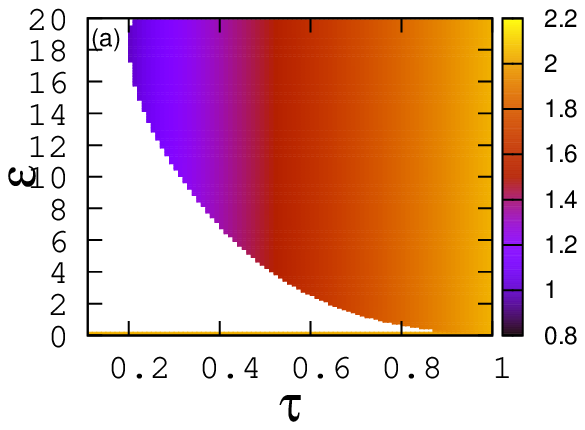}\includegraphics [width = 0.48\columnwidth]{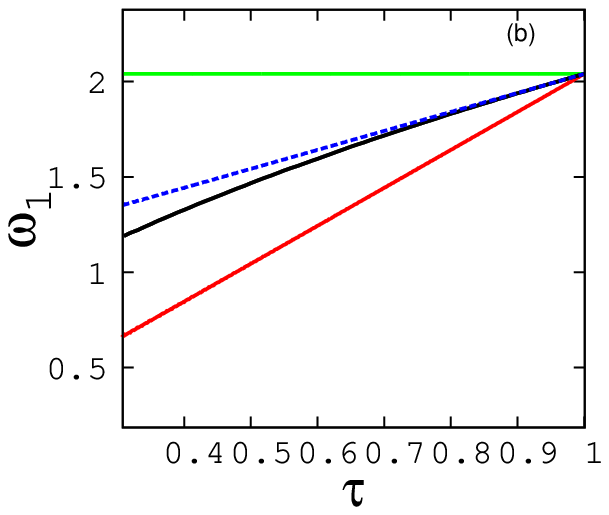}

\caption{(colour online)(a)Contour plot of emergent frequency of two oscillators in $\tau, \epsilon$ plane,(b)plot of frequencies of fast oscillator (green), slow oscillator (red),coupled oscillator (black), mean of frequencies of both oscillator (blue dotted), as $\tau$ varies, at $\epsilon=10$ indicating frequency suppression in the coupled dynamics.}
\label{freqplane}       
\end{figure}

\begin{figure}[H]

  \includegraphics [width = 0.5\columnwidth]{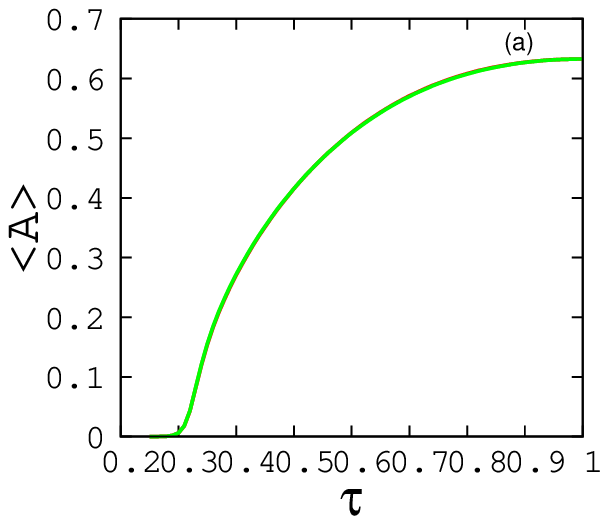}\includegraphics [width = 0.5\columnwidth]{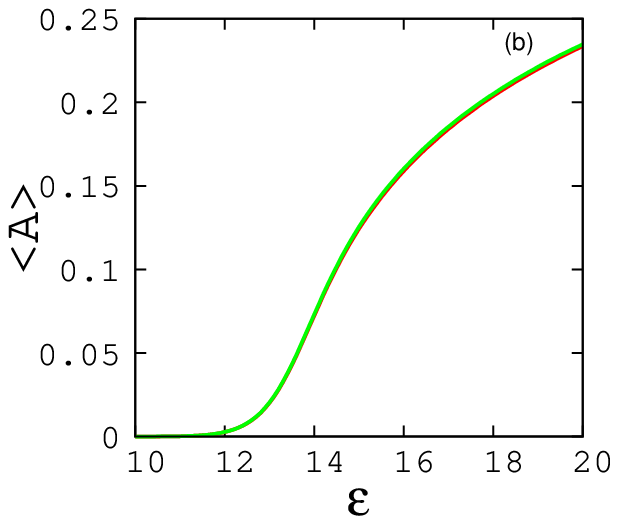}

\caption{(colour online)Average amplitude of slow (red) and fast (green) Landau-Stuart systems $\langle A \rangle$ (a) vs $\tau$ keeping $\epsilon=10$ (b)  vs $\epsilon$ keeping $\tau=0.25$.}
\label{lanamp}       
\end{figure}

We now study the nature of dynamics for parameter values outside the island of AD with strong coupling between systems. When the time scale parameter $ \tau $ is equal to one, i.e. with equal time scales, for sufficient strength of coupling both systems asymptotically will reach the state of complete synchronization. However as $ \tau $ decreases, depending upon the coupling strength the oscillators show different dynamical states. As the time scale mismatch between oscillators increases the two systems fall out of step and soon settle into a state of constant phase relation (Fig~\ref{ssff}), which we find to be a state of frequency synchronization with a phase shift. 
To estimate the phase shift in their states, we calculate the zero crossing time difference ($ t_{k} -t_k' $) between the two oscillators over a sufficient period of time and take the average as $ \phi $. We plot the variation of $\phi$ with varying $\tau$ for a fixed $\epsilon$ shown in Fig~\ref{st}.
The frequency of each oscillator can be calculated from the time series using the relation
\begin{eqnarray}
\omega =\frac {1}{K}\displaystyle\sum_{k=1}^{K}  \frac {2\pi}{(t_ {k+1}-t_k)}
\label{freq}
\end{eqnarray}
where $ t_k $ is the time of the $ k^{th} $ zero crossing point in the time series of the oscillator and K is the total number of intervals used. For sufficiently large $\epsilon$ we find both oscillators settle to the same frequency, resulting in a state of frequency synchronization with a phase shift. This common frequency varies with the changes in values of $\tau$ and $\epsilon$ as shown in the contour plot in the $(\tau,\epsilon)$ plane (Fig~\ref{freqplane}a). Its variation with $\tau$ is compared with the intrinsic frequencies of both oscillators and their average in Fig~\ref{freqplane}b .  As is clear the system with the greater frequency slows down, and the other speeds up. However the frequency of the coupled system is always less than the mean of the frequencies of the uncoupled oscillators, indicating frequency suppression\cite{{ramana00},{ernst91}} (Fig~\ref{freqplane}).
The amplitude of coupled oscillators also depends on the parameters $\tau$ and $\epsilon$ and Fig~\ref{lanamp} shows how average amplitude decreases to zero as amplitude death is reached along both directions of decreasing $\tau$ and $\epsilon$. 
\subsubsection {Multi periodicity under weak coupling}

\begin{figure}[H]

  \includegraphics[width = \columnwidth]{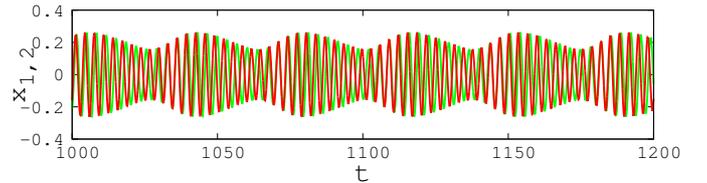}

\caption{(colour online)Time series of the two frequency state for coupled slow (red) and fast (green) Landau-Stuart oscillators for $\tau=0.9 $ and $ \epsilon=0.11 $.}
\label{beat}       
\end{figure}

\begin{figure}[H]

  \includegraphics[width = 0.5\columnwidth]{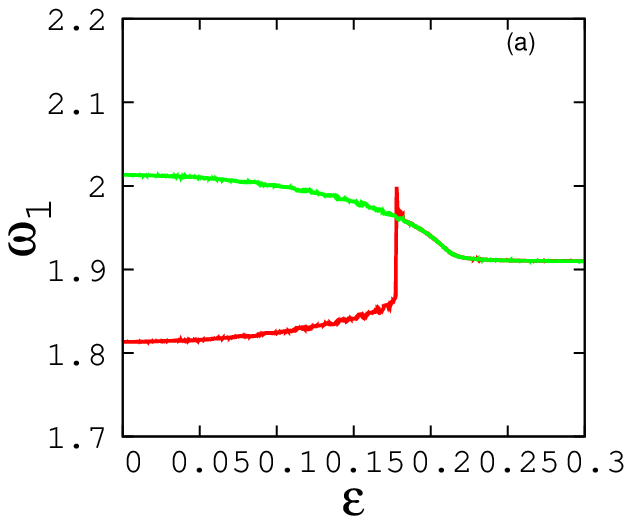}\includegraphics [width = 0.5\columnwidth]{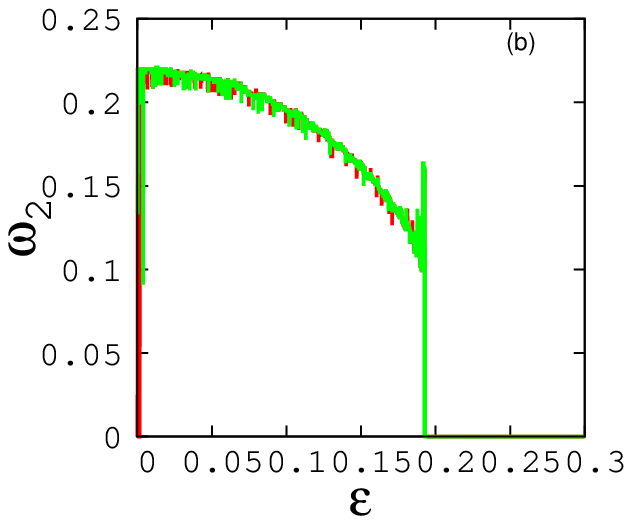}

\caption{(colour online)Variation of frequency with $ \epsilon $ at $ \tau=0.9 $.(a)large frequency of fast (green) and slow (red) systems. (b) small frequency of both systems.}
\label{freq2}       
\end{figure}

We now discuss below the dynamics in the region the AD island for very low coupling strength.
When the coupling strength is very small ($ \epsilon<0.2 $), for a very small mismatch in time scales such as $ \tau=0.9 $ we see each system settles to a two frequency state of oscillation as shown in Fig~\ref{beat}.
Using eqn~(\ref{freq})we calculate the large frequency ($\omega_1$) of each oscillator from the time series data, while the small frequency ($ \omega_2 $) is obtained by taking $ t_k $ as the time of $ k^{th} $ local maximum of all the maxima.
We find that the large frequencies  ($ \omega_1 $) differ, but the small frequencies ($ \omega_2 $) are the same for both the oscillators. As the coupling strength increases the small frequency disappears and the two systems get locked into a state of equal frequency. The variation of $ \omega_1 $ and $ \omega_2 $ as $ \epsilon $ increases is shown in Fig~\ref{freq2}a. Our analysis of the time series using Fast Fourier transform also confirms the above result. 

We repeat the above analysis with periodic R{\"o}ssler system as another example of coupled periodic systems. The equations for two such coupled slow and fast systems are given by
\begin{eqnarray}
 \dot{x}_1&=& \tau(-y_1-z_1) +\tau \epsilon (x_2-x_1)
\nonumber\\
\dot{y}_1&=&\tau(x_1+ay_1) 
 \nonumber\\
\dot{z}_1 &=& \tau(b+z_1(x_1-c))
 \nonumber\\
\dot{x}_2&=&(-y_2-z_2) + \epsilon (x_1-x_2)
 \nonumber\\
\dot{y}_2 &=&(x_2+ay_2) 
 \nonumber\\
\dot{z}_2&=&(b+z_2(x_2-c)) 
\label{prosseqn}
\end{eqnarray}
The intrinsic dynamics is periodic with parameters chosen as a=0.1, b=0.1 and c=4. We observe qualitatively similar results with occurrence of amplitude death, phase locked frequency synchronization with phase shift and two frequency states. The amplitude death region in ($ \tau,\epsilon $) obtained numerically is in good agreement with that from  stability analysis for the synchronized fixed point $ (x^*,y^*,z^*) $ equal to 

\begin{figure}[H]
\begin{center}
  \includegraphics [width = 0.5\columnwidth]{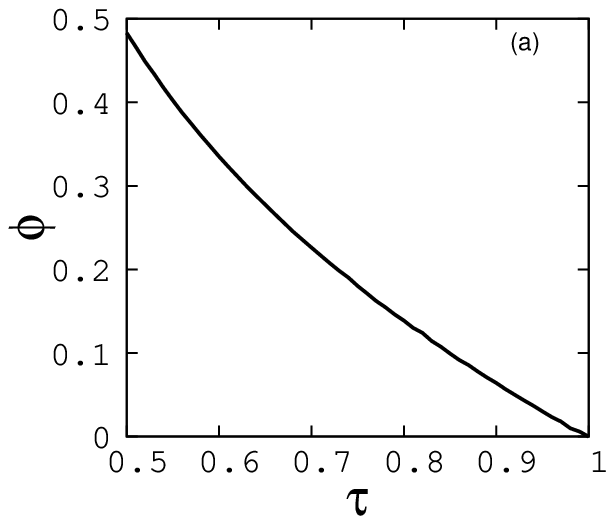}\includegraphics [width = 0.5\columnwidth]{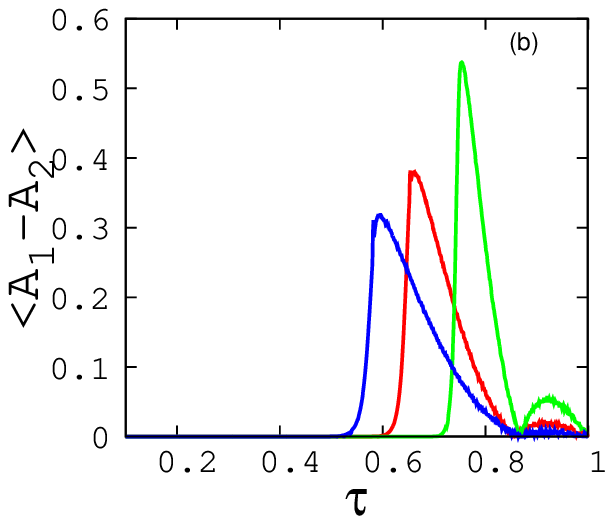}
\end{center}

\caption{(colour online)(a)Phase shift between slow and fast systems $\phi$ vs $\tau$ with $\epsilon$ =1 (b) Average amplitude difference $\langle A_1-A_2 \rangle$ vs $\tau$ for $\epsilon$=0.5(green), $\epsilon $=1 (red), $\epsilon$=1.5(blue), for coupled slow and fast periodic R{\"o}ssler systems.}
\label{pr_phase}       
\end{figure}

$ (\frac {c-\sqrt{c^2-4ab}}{2},\frac {-c+\sqrt{c^2-4ab}}{2a},\frac {c-\sqrt{c^2-4ab}}{2a}) $.
\begin{figure}[H]
\begin{center}
  \includegraphics [width = 0.7\columnwidth]{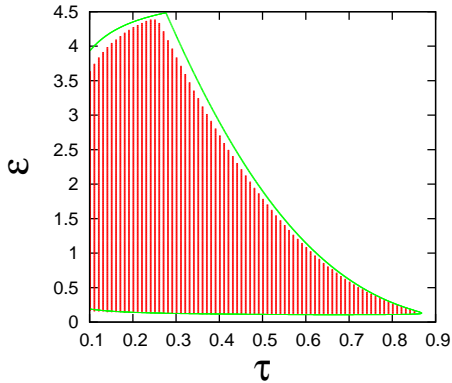}
\end{center}

\caption{(colour online)AD region for coupled slow and fast periodic R{\"o}ssler systems in $\tau$, $\epsilon$ plane.}
\label{pararossler}       
\end{figure}
The phase shift in the frequency synchrnised state and amplitude difference between the oscillators is also observed with the variation of $\tau$ which is shown in the Fig~\ref{pr_phase}.
\subsection{Coupled slow and fast chaotic systems}
We repeat the study for the case of two coupled slow and fast chaotic R{\"o}ssler systems, as in equation~(\ref{prosseqn}), but with parameter values of each system chosen such that their intrinsic dynamics is chaotic (a=0.2, b=0.2, c=5.7).
We also consider coupling two slow and fast Lorenz systems as
\begin{eqnarray}
\dot{x}_1&=&\tau a(y_1-x_1) + \tau\epsilon (x_2-x_1)
\nonumber \\
\dot{y}_1 &=& \tau(x_1(b-z_1)-y_1) 
\nonumber \\
\dot{z}_1 &=& \tau(x_1y_1-cz_1)
\nonumber \\
\dot{x}_2 &=& a(y_2-x_2) + \epsilon (x_1-x_2)
\nonumber \\
\dot{y}_2  &=&x_2(b-z_2)-y_2 
\nonumber \\
\dot{z}_2  &=& x_2y_2-cz_2
\label{lorenzeqn}
\end{eqnarray}
where a=10,b=28,c=8/3. \\

\begin{figure}[H]
\centering
  \includegraphics [width = 0.5\columnwidth]{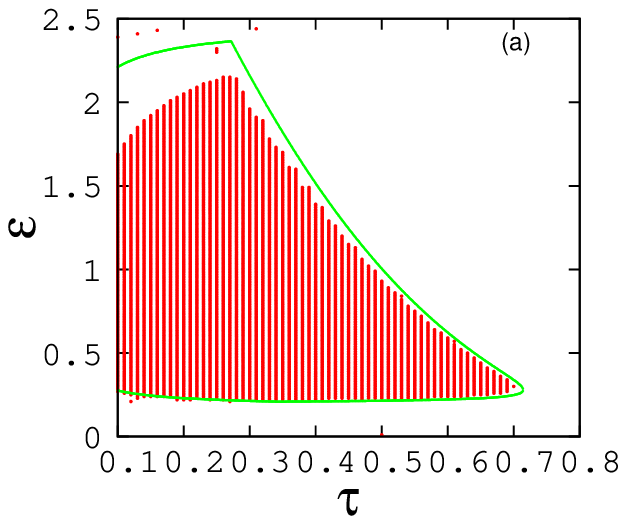}\includegraphics [width = 0.5\columnwidth]{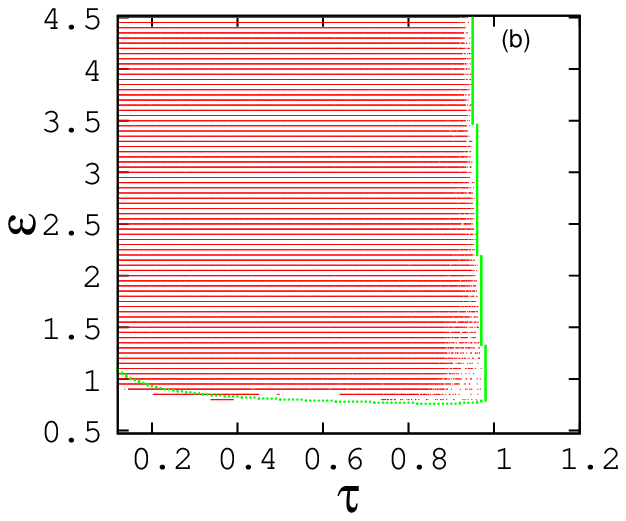}

\caption{(colour online)Parameter plane  ($\tau$, $\epsilon$) showing region of AD. (a) coupled slow and fast chaotic R{\"o}ssler systems (b) coupled slow and fast Lorenz system. }
\label{chrosslerad}       
\end{figure}

\begin{figure}[H]
\centering
\includegraphics [width = 0.9\columnwidth]{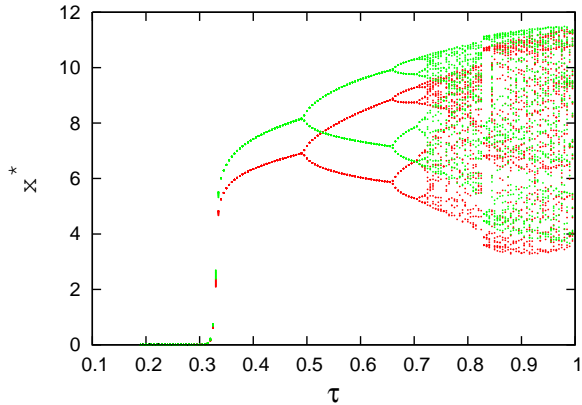}
\caption{\label{rbifurc}(colour online)Bifurcation diagram obtained by plotting the maximum vales of the x variables of the two coupled chaotic R{\"o}ssler systems for $\epsilon=2 $ as $\tau $ is varied.}
\end{figure}

\begin{figure}[H]
\centering

\includegraphics [width = 0.4\columnwidth]{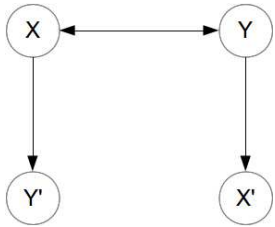}

\caption{\label{sys}Auxiliary systems coupled to slow and fast systems. X-slow system, Y-fast system, X$'$-slow auxiliary system, Y$'$- fast auxiliary system.}
\end{figure}

\begin{figure}[H]
\centering
\begin{center}
\includegraphics [width = 0.55\columnwidth]{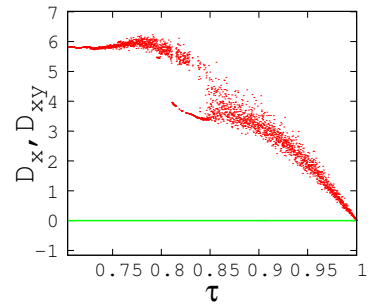}
\end{center}
\caption{\label{ross_gs}(colour online)Plot of the distance between auxiliary systems $D_{x}$or $D_{y}$ (green),  and between main slow and fast systems, $D_{xy}$(red) as $\tau$ is varied.}
\end{figure}

For both the above cases we find that with sufficient strength of coupling and time scale mismatch the systems settle to a state of synchronized fixed point or amplitude death(Fig~\ref{chrosslerad}). The region for which the coupled dynamics of both of the systems goes to amplitude death in the plane ($ \tau,\epsilon $) obtained numerically shows good agreement with the stability analysis. 
We observe that the transition to AD for coupled chaotic R{\"o}ssler systems is through reverse period doublings resulting in periodic dynamics before AD is reached. Even though the amplitudes are different, the bifurcations occur at the same parameter values in both the  systems. The bifurcation diagram corresponding to these transitions as $\tau$ is varied for $ \epsilon =0.9 $, is given in Fig.~\ref{rbifurc}.  After the systems reach periodic state, the average phase shift and average amplitude difference have qualitatively similar behaviour, as in the periodic case described in sec. 2.1. 
However when they remain in the chaotic region for large $\tau$ , we find the systems settle to a state of generalised synchronization. To check this, we attach one slow auxiliary system (X$'$) to the fast system (Y) and one fast auxiliary system (Y$'$) to the slow system (X) unidirectionally, as per the scheme described in \cite{gensyn} for bidirectionally coupled systems (Fig.~\ref{sys}).
 We take the time averaged Euclidian distance between the slow auxiliary systems X and X$'$ as $D_x$ and fast auxiliary systems Y and Y$'$ as $D_y$. For the observed range of $\tau$ shown in the Fig.~\ref{ross_gs}, we observe that $D_x$ and $D_y$ go to zero indicating complete synchronization in the auxiliary systems and generalised synchronization in the main slow and fast systems X and Y (Fig.~\ref{ross_gs}). 

\begin{figure}[H]
\centering
\includegraphics [width = \columnwidth]{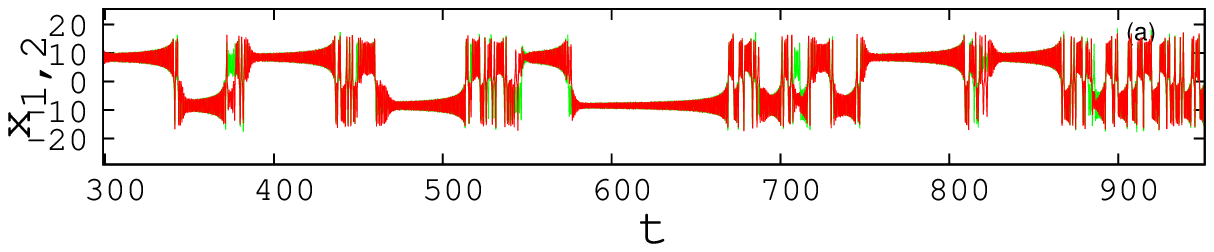}
\includegraphics [width = \columnwidth]{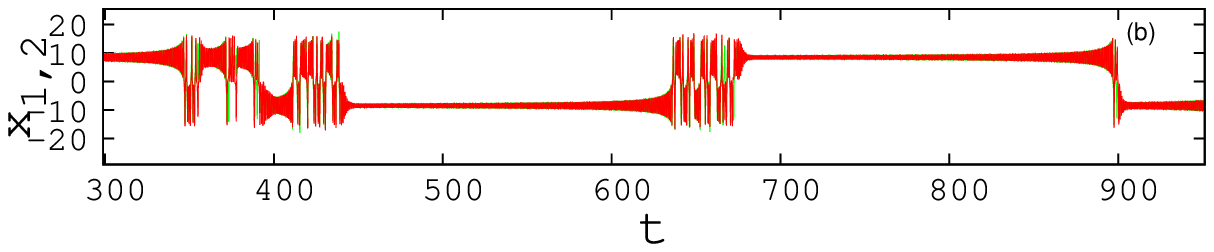}
\includegraphics [width = \columnwidth]{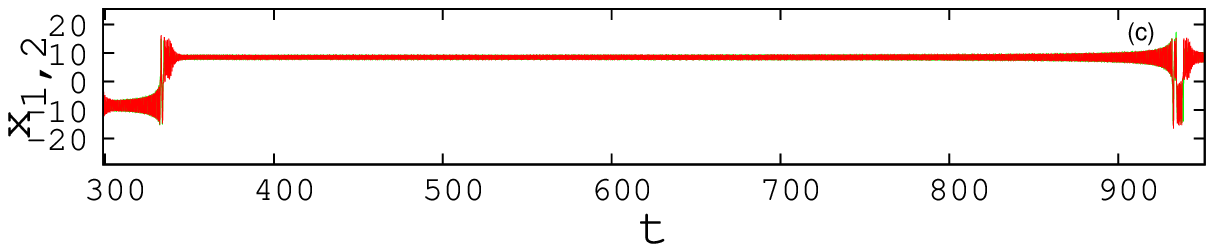}
\caption{\label{lorenz}(colour online)Transition to amplitude death in two coupled chaotic Lorenz systems in ~(\ref{lorenzeqn}). Time series plotted for $\tau=0.949 $ (a), $ 0.948 $(b) and $ 0.947 $ (c) and $ \epsilon = 4$.} 
\end{figure}

However for coupled Lorenz systems the transition to AD is through an intermittency behaviour where the duration of the small amplitude oscillations gets longer as $ \tau $ is decreased. The time series of the coupled Lorenz systems are plotted for increasing values of $ \tau $ with $ \epsilon =4.0 $ in Fig.~\ref{lorenz} which indicates this intermittency route to AD.

\section{Coupled Ocean Atmosphere model}

As an important application of the phenomena introduced in the previous sections, we consider the coupled ocean atmosphere model used in climate studies\cite{{siq12},{sergei14}}.  In this context, it is usual to consider low dimensional Lorenz system as the basic dynamics and couple two versions of the same , one with fast and other with slow time scales, to model the interactive dynamics of a fast oscillating atmosphere and slow-fluctuating ocean.  
The equations representing coupled convective dynamics studied earlier are given below \cite{sergei14}.
\begin{eqnarray}
\dot{x}_1&=&\tau a(y_1-x_1) - \epsilon x_2
\nonumber \\
\dot{y}_1 &=& \tau(x_1(b-z_1)-y_1) +\epsilon y_2
\nonumber \\
\dot{z}_1 &=& \tau(x_1y_1-cz_1)-\epsilon z_2
\nonumber \\
\dot{x}_2 &=& a(y_2-x_2) - \epsilon x_1
\nonumber \\
\dot{y}_2  &=&x_2(b-z_2)-y_2 +\epsilon y_1
\nonumber \\
\dot{z}_2  &=& x_2y_2-cz_2+\epsilon z_1
\label{lorenzclimate}
\end{eqnarray}
where a=10,b=28,c=8/3 and $\tau$ is the slow time-scale parameter.\\
We revisit this model from the point of view of the above analysis and report the interesting dynamics resulting in periodic and steady state convection due to the interaction. Unlike the previous cases, we observe dissimilar attractors for slow and fast systems, multi stable periodic states and oscillation death etc.

\subsection{Oscillation death}
We numerically analyse the coupled model in this case and find for a certain region in the parameter plane $(\tau,\epsilon)$ the coupled dynamics go to oscillation death (OD) i.e the two systems goes to two different fixed points(fig~\ref{lcod}). 
\begin{figure}[H]
\centering
\includegraphics [width = \columnwidth]{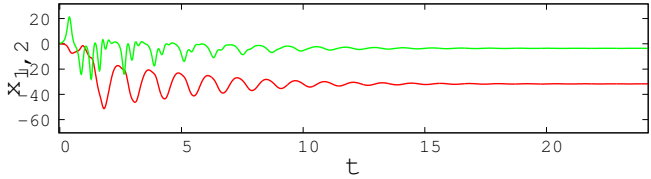}
\caption{\label{lcod}(colour online)Time series of coupled ocean-atomosphere model at oscillation death at $\tau=0.15, \epsilon=3 $. } 
\label{lcod}
\end{figure}
This region of OD is shown in (Fig.~\ref{lorenzclimate}). In the parameter plane the region above the upper boundary corresponds to unstable behaviour, while that below shows periodic dynamics and multistable states.
\begin{figure}[H]
\centering
\includegraphics [width = \columnwidth]{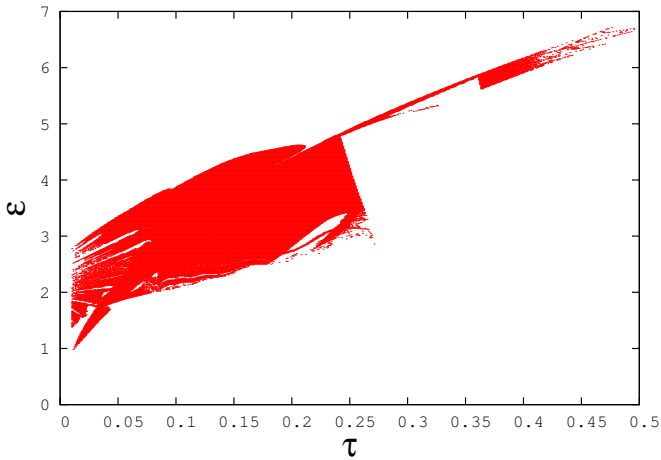}
\caption{\label{lorenzclimate}(colour online)Region of oscillation death in coupled ocean-atmosphere system in $\tau, \epsilon $ plane.} 
\end{figure}

\subsection{Periodic oscillations and Multistable states}
\begin{figure}[H]
\centering
\includegraphics [width = 0.5\columnwidth]{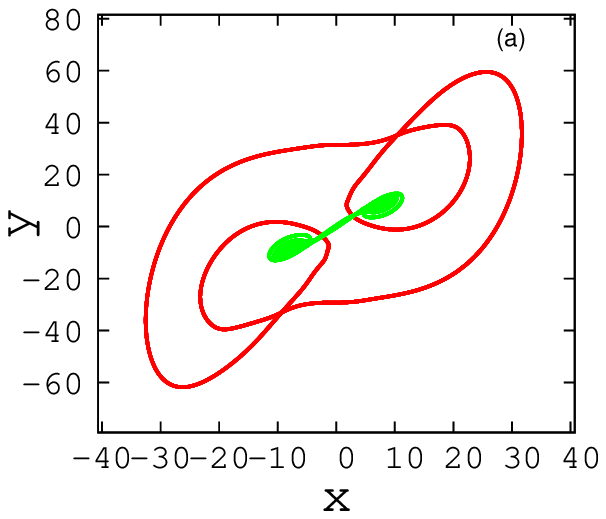}\includegraphics [width = 0.5\columnwidth]{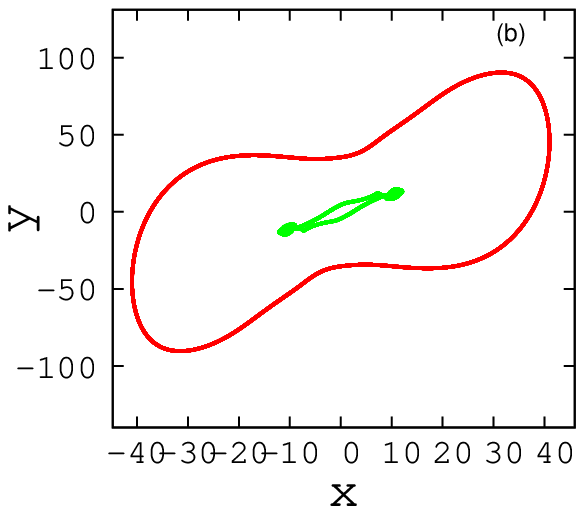}

\includegraphics [width = 0.5\columnwidth]{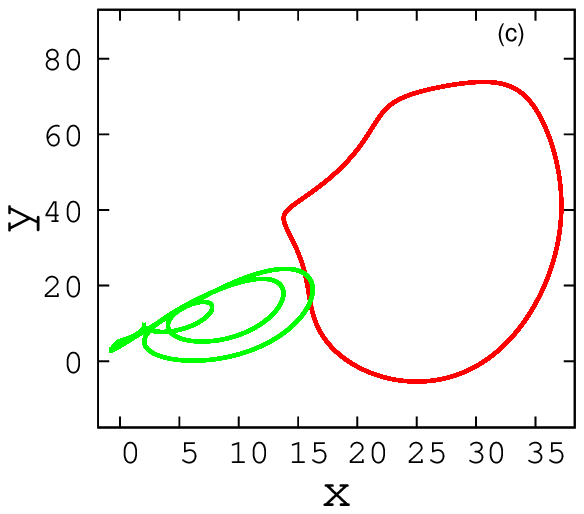}\includegraphics [width = 0.5\columnwidth]{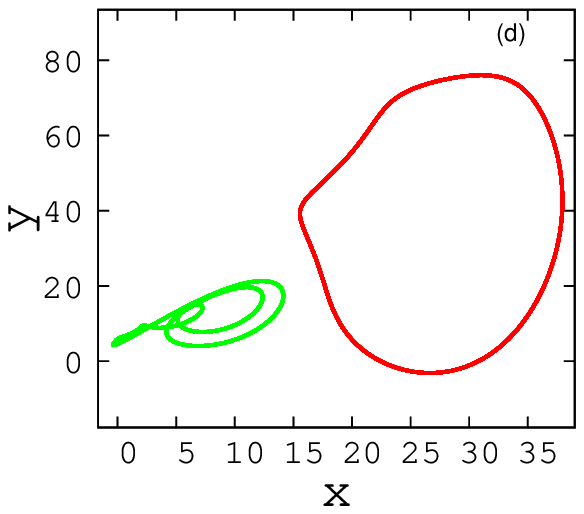}
\caption{\label{lorenzclimateod}(colour online)Attractors of coupled ocean-atmosphere model before reaching OD, for $\tau=0.1$, $\epsilon=0.8$ (a),$\epsilon=1.5$(b),$\epsilon=1.92$ (c) and $\epsilon=2.02$ (d) in X-Y plane.} 
\end{figure}

\begin{figure}[H]
\centering
  \includegraphics [width = 0.5\columnwidth]{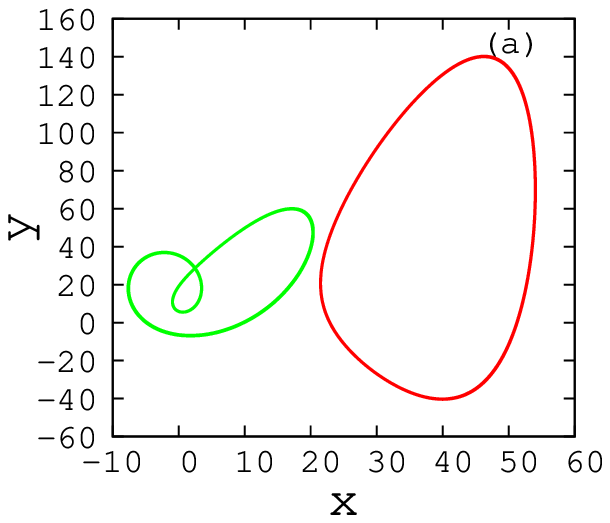}\includegraphics [width = 0.5\columnwidth]{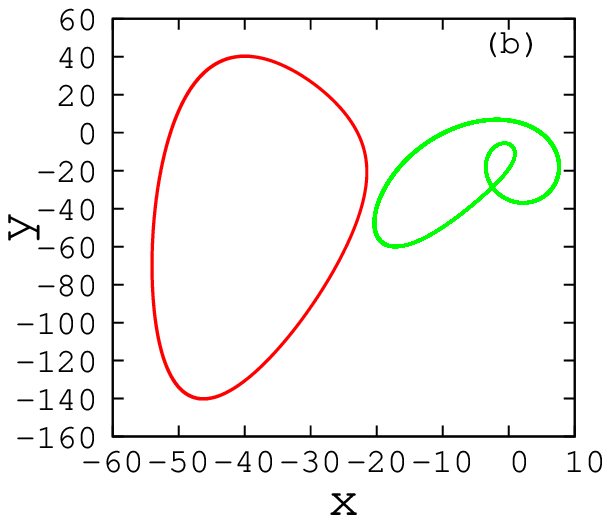}
\includegraphics [width = \columnwidth]{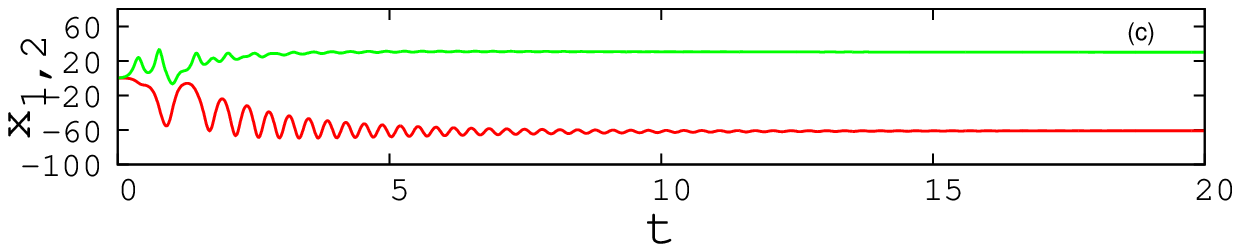}
\includegraphics [width = \columnwidth]{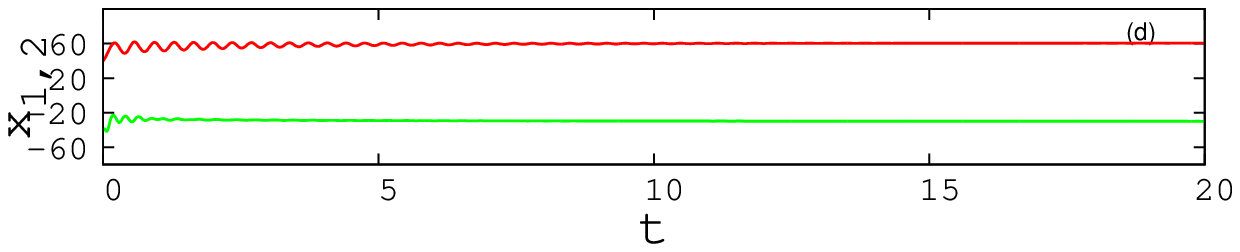}
\caption{(colour online)Periodic states(a),(b) and Oscillation death states(c),(d) of coupled ocean-atmosphere model at $\tau=0.312 $, $\epsilon=5.4$ for different initial conditions.}
\label{lorenzphase}       
\end{figure}

 For large values of timescale smismatch as we increase the coupling between the oscillators we see periodic behaviour in this model. This is clear from Fig.~\ref{lorenzclimateod} where for values of $\epsilon$ and $\tau$ below the OD region, the phase space plots in XY plane are shown. 
We also find multistable states below the narrow strip of OD in the parameter plane. Here for the same parameter value, oscillation death states and periodic states are found to occur for different initial conditions. We analyse this, by keeping $(y_1,z_1,y_2,z_2)$ as $(0.3,0.4,0.5,0.6)$ and varying $(x_1,x_2)$. Thus for $(x_1,x_2)$= $(0.8,0.34),(0.1,0.5),\\(0.4,0.5),(-40,40)$ the different possible states for the same value of $(\tau,\epsilon)$ are shown in Fig.~\ref{lorenzphase}a,b,c and d respectively. It is clear that the systems settle to two types of oscillatory states and two types of OD states indicating multistability.

\begin{figure}[H]
\centering
\includegraphics [width = \columnwidth]{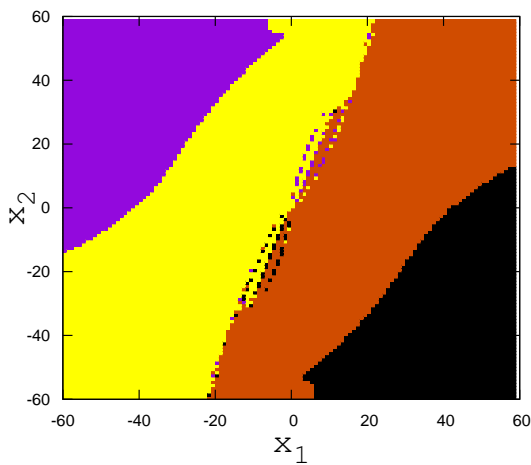}
\caption{(colour online)Basin structure for the multistable state in coupled ocean-atmosphere system. The region in blue/black form the basin for the oscillation death states shown in Fig~\ref{lorenzphase}(c)/(d)), while red and yellow region correspond to periodic oscillation shown in Fig~\ref{lorenzphase} (a) and (b) respectively. }
\label{lorenzbasin}       
\end{figure}

The occurence of multistability is associated with different basins of attraction in the initial value space of the coupled system. We study this structure by scanning the $(x_1,x_2)$ plane between $(-60,60)$ keeping $(y_1,z_1,y_2,z_2)$ same as in Fig.~\ref{lorenzphase} and identify regions of different dynamics. The corresponding basin structure thus obtained is shown in Fig.~\ref{lorenzbasin}. 
\section{Conclusion}
The study presented here illustrates that in general, mismatch in time scales of different interacting units can affect the group performance of the coupled system leading to suppression of dynamics. In the context of coupled systems, we report this as another mechanism that can lead to amplitude death. Taking two coupled periodic systems like Landau-Stuart and R{\"o}ssler with differing time scales, we show that the systems get locked to a state of frequency synchronization with a phase shift and undergo transition to amplitude death as the strength of interaction or mismatch in time scale increases. We fiurther extend the study to coupled chaotic systems and study the transition to amplitude death in them.

We also show a interesting application of the analysis to the coupled climate model represented by two coupled Lorenz systems with differing time scales. In this case, we observe the suppression of dynamics resulting in a state of oscillation death that can correspond to steady state convection of differing amplitudes in the atmosphere and ocean. In addition, there can be coexisting multi-stable states of periodic and steady states. In the context of climate studies, such multistability with the consequent complex basin structure shown would indicate added unpredictability, even when the dynamics is periodic or steady state convection.

The concepts introduced are quite general and can be applied to any type of interacting systems evolving at different time scales. In such cases how the connections can affect the collective behavior and how they can be controlled or prevented are relevant questions to be addressed. Studies in this direction are currently in progress and the results will be reported elsewhere.\\


One of the authors (K.G) would like to thank University Grants Commission, New Delhi, India for financial support.



\end{document}